\documentclass[pra,twocolumn,aps,showpacs,superscriptaddress]{revtex4-1}
\usepackage{epsfig,graphicx,tabularx}
\usepackage{psfrag}
\usepackage{graphicx}
\usepackage{graphics}
\usepackage{epsfig}
\usepackage{bm}
\usepackage{verbatim,color,ulem}
\usepackage{hyperref, comment}
\usepackage{amssymb}
\usepackage{physics}

\newcommand{\beq}{\begin{equation}}
\newcommand{\eeq}{\end{equation}}
\newcommand{\beqa}{\begin{eqnarray}}
\newcommand{\eeqa}{\end{eqnarray}}

\newcommand{\saw}{\textcolor{black}}
\newcommand{\cred}{\color{black}}

\begin{document}

\title{Finite-size effects in a bosonic Josephson junction}

\author{Sandro Wimberger}
\affiliation{Dipartimento di Scienze Matematiche, Fisiche ed Informatiche, 
Universit\`a di Parma, Parco Area delle Scienze 7/A, 43124 Parma, Italy}
\affiliation{INFN - Sezione di Milano-Bicocca, gruppo collegato di Parma, 
Parco Area delle Scienze 7/A, 43124 Parma, Italy}

\author{Gabriele Manganelli}
\affiliation{Dipartimento di Fisica e Astronomia 'Galileo Galilei', 
Universit\`a di Padova, via Marzolo 8, 35131 Padova, Italy}
\affiliation{Scuola Galileiana di Studi Superiori, 
Universit\`a di Padova, \\ via San Massimo 33, 35129 Padova, Italy}

\author{Alberto Brollo}
\affiliation{Dipartimento di Fisica e Astronomia 'Galileo Galilei',  
Universit\`a di Padova, via Marzolo 8, 35131 Padova, Italy}

\author{Luca Salasnich}
\affiliation{Dipartimento di Fisica e Astronomia 'Galileo Galilei', 
Universit\`a di Padova, via Marzolo 8, 35131 Padova, Italy}
\affiliation{Padua Quantum Technologies Research Center, 
Universit\`a di Padova, \\ via Gradenigo 6/b, 35131 Padova, Italy}
\affiliation{INFN - Sezione di Padova, via Marzolo 8, 35131 Padova, Italy}
\affiliation{CNR-INO, via Nello Carrara 1, 50019 Sesto Fiorentino, Italy}

\begin{abstract}
We investigate finite-size quantum effects in the 
dynamics of $N$ bosonic particles which are tunneling between two sites 
adopting the two-site Bose-Hubbard model. 
By using time-dependent atomic coherent states (ACS) 
we extend the standard mean-field  
equations of this bosonic Josephson junction, which are 
based on time-dependent Glauber coherent states. In this way 
we find $1/N$ corrections to familiar mean-field (MF) results: 
the frequency of macroscopic oscillation between the two sites, 
the critical parameter for the dynamical macroscopic quantum 
self trapping (MQST), and the attractive critical interaction strength  
for the spontaneous symmetry breaking (SSB) of the ground state. 
To validate our analytical results we perform {\cred numerical} 
simulations of the quantum dynamics. In the case of Josephson 
oscillations around a balanced configuration 
we find that also for a few atoms the numerical results 
are in good agreement with the predictions of time-dependent 
ACS variational approach, provided that the time evolution 
is not too long. Also the {\cred numerical} results of 
SSB are better reproduced by the ACS approach with respect to the MF one. 
Instead the onset of MQST is correctly reproduced by ACS theory 
only in the large $N$ regime and, for this phenomenon, 
the $1/N$ correction to the MF formula is not reliable. 
\end{abstract}

\pacs{03.75.Lm; 74.50.+r}

\maketitle

\section{Introduction}

The Josephson junction is a quantum mechanical device 
made of two superconductors, or two superfluids, separated by a 
tunneling barrier \cite{josephson1962}. The Josephson junction can give rise 
to the direct-current (DC) Josephson effect, where a supercurrent flows 
indefinitely long across the barrier, but also to the alternate-current (AC) 
Josephson effect, where due to an energy difference the supercurrent oscillates 
periodically across the barrier \cite{barone1982}. 
The superconducting quantum interference devices (SQUIDs), which are very 
sensitive magnetometers based on superconducting Josephson junctions, 
are widely used in science and engineering \cite{vari2017}. Moreover, 
{\cred Josephson junctions are now used to realize qubits 
(see, for instance, \cite{quantum1,quantum2}).} 

The achievement of Bose-Einstein condensation with 
ultracold and dilute alkali-metal atoms \cite{bec1995} has renewed and 
increased the interest on macroscopic quantum phenomena and, 
in particular, on the Josephson effect \cite{bloch2008}. Indeed, contrary to 
the case of superconducting Josephson junctions, with 
atomic Josephson junctions it is possible to have a large population 
imbalance with the appearance of the self-trapping 
phenomenon \cite{smerzi1997}. A direct experimental observation 
of {\cred tunneling} and nonlinear self-trapping in a single 
bosonic Josephson junction was made in 2005 with $^{87}$Rb 
atoms \cite{albiez2005}. More recently, in 2015, Josephson effect 
has been detected in fermionic superfluids across the BEC-BCS crossover 
with $^6$Li atoms \cite{valtolina2015}. 

The fully quantum behavior of Josephson junctions is usually described 
by using the phase model \cite{leggett1991}, which is based on the 
quantum commutation rule \cite{commutation} between the number 
operator ${\hat N}$ and the phase angle operator ${\hat \phi}$. 
Within this model it has been found that 
quantum fluctuations renormalize the mean-field Josephson oscillation 
\cite{smerzi2000,anglin2001,ferrini2008}. However, the phase angle operator 
${\hat \phi}$ is not Hermitian, the exponential phase operator 
$e^{i{\hat \phi}}$ is not unitary, and their naive application can 
give rise to wrong results. Despite such problems, 
the phase model is considered a good starting point in many theoretical 
studies  of Josephson junctions, because the phase-number commutation rule 
is approximately correct for systems with a large number 
of condensed electronic Cooper-pairs or bosonic atoms \cite{anglin2001}. 

In this paper we study finite-size quantum effects in a 
Josephson junction avoiding the use of the phase operator. 
The standard mean-field theory is based on the 
Glauber coherent state $|CS\rangle$ 
which however is not eigenstate of the total number 
operator \cite{glauber1963}. 
Here we adopt the atomic coherent state $|ACS\rangle$, 
which is instead eigenstate of the total number operator, 
and it reduces to the Glauber coherent state only in the limit of 
a large number $N$ of bosons 
\cite{arecchi1972,gilmore1990,walls1997,penna2005,penna2006,penna2008}. 
We prove that the frequency of macroscopic oscillation of bosons 
between the two sites is given by $\sqrt{J^2+NUJ(1-1/N)}/\hbar$, 
where $J$ is the tunneling energy, $U$ is the on-site interaction 
energy. Remarkably, for very large number $N$ of bosons this formula becomes 
the familiar mean-field one $\sqrt{J^2+NUJ}/\hbar$. 
We find similar corrections for the critical 
strength of the dynamical self-trapping and for the critical strength 
of the population-imbalance symmetry breaking of the ground state. 
Once again in these cases the standard mean-field 
results are retrieved in the limit of a large number $N$ of bosons. 
In the last part of the paper we compare the ACS theory with 
{\cred numerical} simulations. In the case of Josephson oscillations 
we find a very good agreement between ACS theory and numerical results 
also for a small number $N$ of bosons. For the ACS 
critical interaction strength of the semiclassical 
spontaneous symmetry breaking of the ground state 
we obtain a reasonable agreement with the {\cred numerical} results. 
Instead, for the phenomenon of self-trapping, our {\cred numerical} quantum 
simulations suggest that the $1/N$ corrections predicted by the ACS theory 
are not reliable. We attribute this discrepancy to the increased 
importance of quantum fluctuations and stronger many-body 
correlation in the so-called Fock regime, see e.g. \cite{leggett2001}.

\section{Two-site model}

The macroscopic quantum tunneling of bosonic particles or Cooper 
pairs in a Josephson junction made of two superfluids or two 
superconductors separated by a potential barrier can {\cred be} described 
within a second-quantization formalism, see for instance \cite{lewenstein}. 
The simplest quantum Hamiltonian of a system made of bosonic particles 
which are tunneling between two sites ($j=1,2$) is given by 
\beq 
{\hat H} =  - {J} \left( {\hat a}_1^+ {\hat a}_2 + {\hat a}_2^+ 
{\hat a}_1 \right) + {U} \sum_{j=1,2} {\hat N}_j ({\hat N}_j-1) \; , 
\label{ham}
\eeq
where ${\hat a}_j$ and ${\hat a}_j^+$ are the dimensionless ladder operators 
which, respectively, destroy and create a boson in the $j$ site, 
${\hat N}_j={\hat a}_j^+{\hat a}_j$ is the number operator 
of bosons in the $j$ site. 
$U$ is the on-site interaction strength of particles and 
{\cred $J>0$} is the tunneling energy, both measured in units of the reduced 
Planck constant $\hbar$. 
Eq. (\ref{ham}) is the so-called two-site Bose-Hubbard Hamiltonian. 
We also introduce the total number operator 
\beq
{\hat N} = {\hat N}_1 + {\hat N}_2 \; . 
\label{ntot}
\eeq

The time evolution of a generic quantum state $|\psi(t)\rangle$ 
of our system described by the Hamiltonian (\ref{ham}) 
is then given by the Schr\"odinger equation 
\beq 
i \hbar {\partial \over \partial t} |\psi(t)\rangle = {\hat H} 
|\psi(t)\rangle \; . 
\label{rompino}
\eeq
Quite remarkably, this time-evolution equation can be derived 
by extremizing the following action 
\beq 
S = \int dt \, \langle \psi(t) | \left( i \hbar {\partial \over \partial t} - 
{\hat H} \right) |\psi(t) \rangle \; ,  
\label{action}
\eeq
characterized by the Lagrangian 
\beq 
L =  i\hbar \langle\psi(t)|{\partial \over \partial t}|\psi(t)\rangle  
- \langle\psi(t)|{\hat H} |\psi(t) \rangle \; . 
\label{lag}
\eeq
{\cred Clearly, Eqs. (\ref{rompino}--\ref{lag}) hold for any quantum system.} 

\section{Standard mean-field dynamics} 

The familiar mean-field dynamics of the bosonic Josephson junction 
can be obtained with a specific choice for the quantum state $|\psi(t)\rangle$, 
namely \cite{penna1998}
\beq 
|\psi(t)\rangle = |CS(t)\rangle \; , 
\eeq
where 
\beq 
|CS(t)\rangle = |\alpha_1(t) \rangle \otimes 
|\alpha_2(t) \rangle \; 
\label{cs}
\eeq
is the tensor product of Glauber coherent states $|\alpha_j(t)\rangle$,  
defined as 
\beq 
|\alpha_j(t) \rangle = e^{-{1\over 2}|\alpha_j(t)|^2} \ 
e^{\alpha_j(t) {\hat a}_j^+} |0\rangle 
\label{new}
\eeq
with $|0\rangle$ the vacuum state, and such that 
\beq
{\hat a}_j|\alpha_j(t)\rangle = \alpha_j(t)|\alpha_j(t)\rangle \; .  
\eeq
Thus, $|\alpha_j(t)\rangle$ is the eigenstate of the annihilation operator 
${\hat a}_j$ with eigenvalue $\alpha_j(t)$ \cite{glauber1963}. 
The complex 
eigenvalue $\alpha_j(t)$ can be written as 
\beq 
\alpha_j(t) = \sqrt{N_j(t)} \, e^{i\phi_j(t)} \; , 
\eeq
with $N_j(t)=\langle\alpha_j(t)|{\hat N}_j|\alpha_j(t)\rangle$ 
the average number of bosons in the site $j$ at 
time $t$ and $\phi_j(t)$ the corresponding phase angle at the same time $t$. 

Adopting the coherent state (\ref{cs}) with Eq. (\ref{new}) 
the Lagrangian (\ref{lag}) becomes 
\beqa 
L_{CS} &=&  i\hbar \langle CS(t)|{\partial \over \partial t}|CS(t)\rangle  
- \langle CS(t)|{\hat H} |CS(t) \rangle  
\nonumber 
\\
&=&  N \hbar \, z {\dot \phi} - {U N^2\over 2} z^2 
+ J N \sqrt{1-z^2} \, \cos{(\phi)} \; ,  
\label{golden}
\eeqa
where the dot means the derivative with respect to time $t$, 
\beq 
N=N_1(t)+N_2(t)
\eeq
is the average total number of bosons (that is a constant 
of motion), 
\beq 
\phi(t)=\phi_2(t)-\phi_1(t)
\eeq
is the relative phase, and 
\beq 
z(t) = {N_1(t) - N_2(t) \over N}
\eeq
is the population imbalance. The last term in the Lagrangian (\ref{golden})
is the one which makes possible 
the periodic oscillation of a macroscopic number of particles 
between the two sites. 

In the Lagrangian $L_{CS}(\phi,z)$ of Eq. (\ref{golden}) 
the dynamical variables $\phi(t)$ 
and $z(t)$ are the generalized Lagrangian coordinates 
(see, for instance, \cite{penna2000}). 
The extremization of the action (\ref{action}) with the Lagrangian 
(\ref{golden}) gives rise to the Euler-Lagrange equations 
\beqa
{\partial {L}_{CS}\over \partial \phi} - {d\over dt}
{\partial {L}_{CS}\over \partial {\dot \phi}} = 0 \; , 
\\
{\partial {L}_{CS}\over \partial z} - {d\over dt}
{\partial {L}_{CS}\over \partial {\dot z}} = 0 \; ,
\eeqa
which, explicitly, become
\beqa 
{\dot \phi} &=& J {z\over \sqrt{1-z^2}} \cos{(\phi)} + U N z \; ,
\label{starting1}
\\
{\dot z} &=& - J \sqrt{1-z^2} \sin{(\phi)} \; .  
\label{starting2}
\eeqa
These equations describe the mean-field dynamics of the macroscopic 
quantum tunneling in a Josephson junction, 
where $\phi(t)$ is the relative phase angle of the 
complex field of the superfluid (or superconductor) 
between the two junctions at time $t$ 
and $z(t)$ is the corresponding relative population imbalance 
of the Bose condensed particles (or Cooper pairs). 

Assuming that both $\phi(t)$ and $z(t)$ are small, 
i.e. $|\phi(t)|\ll 1$ and $|z(t)|\ll 1$, 
the Lagrangian (\ref{golden}) can be approximated as 
\beq 
{L}_{CS}^{(2)} = N \hbar\, z {\dot \phi} - 
{JN\over 2} \phi^2 - {(JN + U N^2)\over 2} z^2 \; ,  
\label{quadratic}
\eeq 
removing a constant term. The Euler-Lagrange equations of this 
quadratic Lagrangian are the linearized Josephson-junction equations
\beqa 
\hbar \, \dot{\phi} &=& (J+UN) z  \; , 
\label{linear-jo1}
\\
\hbar \, \dot{z} &=& - J\phi \; ,  
\label{linear-jo2}
\eeqa
which can be rewritten as a single equation for the 
harmonic oscillation of $\phi(t)$ and the harmonic 
oscillation of $z(t)$, given by 
\beqa 
\ddot{\phi} + \Omega^2 \ \phi = 0 \; ,
\\
\ddot{z} + \Omega^2 \ z = 0 \; , 
\eeqa
both with frequency 
\beq 
\Omega = {1\over \hbar} \sqrt{J^2 + N U J} \; , 
\label{frequenza-vera-mf}
\eeq
that is the familiar mean-field frequency of macroscopic quantum 
oscillation in terms of tunneling energy {\cred $J>0$}, 
interaction strength $U$, and number $N$ of particles \cite{smerzi1997}. 

It is straightforward to find that 
the conserved energy of the mean-field system 
described by Eqs. (\ref{starting1}) and (\ref{starting2}) is given by 
\beq
{E}_{CS} = {U N^2\over 2} z^2 - J N \sqrt{1-z^2} \, \cos{(\phi)} \; . 
\label{ham-cs}	
\eeq
If the condition
\beq
{E}_{CS} (z(0), \phi(0) )> {\cred E_{CS}} (0, \pi) 
\label{trappingcondition}
\eeq
is satisfied then $ \langle z \rangle \neq 0$ since $z(t)$ cannot become zero 
during an oscillation cycle. This situation is known as macroscopic quantum 
self trapping (MQST) {\cred \cite{smerzi1997,rag1999,ashhab2002}.} 
Introducing the dimensionless strength 
\beq 
\Lambda= {NU \over J} \; , 
\eeq
the expression (\ref{ham-cs}) and the trapping condition 
(\ref{trappingcondition}) give
\beq
\Lambda_{MQST} = { {1+\sqrt{1-z^2(0)} \cos(\phi (0))} \over z(0)^2/2}
\label{mqst}
\eeq
for the critical value of $\Lambda$ above which the self trapping occurs. 
Indeed, 
\beq 
\Lambda> \Lambda_{MQST}
\eeq
is the familiar mean field condition to achieve MQST 
in BECs \cite{smerzi1997}. {\cred We stress that MQST condition 
crucially depend on the specific initial conditions $\phi(0)$ and $z(0)$.}  

Let us study the stationary solutions of (\ref{golden}). From the system 
of Eqs. (\ref{starting1}) and (\ref{starting2}) we obtain the symmetric 
solutions 
\beqa
(\tilde{z}_{-}, \tilde{\phi})=(0, 2n \pi )
\label{symmetricplus} \\
(\tilde{z}_{+}, \tilde{\phi})=(0, (2n+1)\pi ) 
\label{symmetricminus}
\eeqa
with $n \in \mathbb{Z}$ ,respectively with energies $\tilde{E}_{-}=-JN$ 
and $\tilde{E}_{+}=JN$. Due to the nonlinear interaction there are degenerate 
{\cred ground-state} solutions that break the z-symmetry
\beqa
z_{\pm}&=&\pm \sqrt{1-{{1} \over {\Lambda^2}}   }
\label{symmetrybroken} \\
\phi_n &=& {\cred 2\pi n} 
\eeqa
where $n \in \mathbb{Z}$. These solutions give a minimum of the 
energy with $\phi=0$ only for $\Lambda=UN/J <0$. Thus, the spontaneous 
symmetry breaking (SSB) of the balanced ground state ($z=0$, $\phi=0$) 
appears at the critical dimensionless strength  
\beq 
\Lambda_{SSB} = - 1  \; . 
\label{ssb-cs}
\eeq
In other words, for $\Lambda=UN/J < \Lambda_{SSB}=-1$ the population imbalance 
$z$ of the ground state of our bosonic system becomes different from zero. 

\section{Finite-size effects}

Different results are obtained 
by choosing another quantum state $|\psi(t)\rangle$ 
in Eqs. (\ref{action}) and (\ref{lag}). In this section, our 
choice for the quantum state $|\psi(t)\rangle$ is 
\beq 
|\psi(t)\rangle = |ACS(t)\rangle \; , 
\eeq
where 
\beq 
|ACS(t)\rangle = 
{\left( \sqrt{1+z(t)\over 2} 
{\hat a}_1^+ + \sqrt{1-z(t)\over 2} 
\, e^{-i\phi(t)}{\hat a}_2^+ \right)^N \over \sqrt{N!}} |0\rangle  \; 
\label{acs}
\eeq
is the atomic coherent state \cite{arecchi1972}, 
also called SU(2) coherent state or Bloch state or angular momentum 
coherent state \cite{gilmore1990}, with $|0\rangle$ the vacuum state. 
This atomic coherent state depends on two dynamical variables 
$\phi(t)$ and $z(t)$ which, as we shall show, can be again interpreted 
as relative phase and population imbalance of the Josephson 
junction \cite{walls1997,penna2005,penna2006,penna2008,
trimborn2008,trimborn2009}.

Contrary to the Glauber coherent state $|CS(t)\rangle$ of Eq. (\ref{cs}), 
the atomic coherent state of Eq. (\ref{acs}) is an eigenstate of the total 
number operator (\ref{ntot}), i.e. 
\beq 
{\hat N} |ACS(t)\rangle = N |ACS(t) \rangle \; . 
\eeq
Moreover, {\cred the averages calculated with the}  
atomic coherent state $|ACS(t)\rangle$ 
{\cred become equal to the ones performed with} 
the Glauber coherent state $|CS(t)\rangle$ only 
in the regime $N\gg 1$ 
\cite{arecchi1972,gilmore1990,walls1997,penna2005,penna2006,penna2008,
trimborn2008,trimborn2009}. 

Adopting the atomic coherent state (\ref{acs}) 
the Lagrangian (\ref{lag}) becomes 
\begin{widetext}
\beqa 
L_{ACS} &=&  i\hbar \langle ACS(t)|{\partial \over \partial t}|ACS(t)\rangle  
- \langle ACS(t)|{\hat H} |ACS(t) \rangle  
\nonumber 
\\
&=&  N \hbar \, z {\dot \phi} - {U N^2\over 2} \left(1 - {1\over N}\right) 
z^2 + J N \sqrt{1-z^2} \, \cos{(\phi)} \; .  
\label{boh}
\eeqa
\end{widetext}
Comparing this expression with the Lagrangian of the Glauber 
coherent state, Eq. (\ref{golden}), one immediately 
observes that the two Lagrangians become equal under the condition 
$N\gg 1$. Moreover, the former is obtained from the latter with the 
formal substitution $U\to U(1-1/N)$. 
In other words, the term $(1-1/N)$ takes into account 
few-body effects, which become negligible only for $N\gg 1$.  

It is immediate to write down the corresponding Josephson equations 
\beqa 
{\dot \phi} &=& J {z\over \sqrt{1-z^2}} \cos{(\phi)} + U N 
\left( 1-{1\over N} \right) z \; ,
\label{b1}
\\
{\dot z} &=& - J \sqrt{1-z^2} \sin{(\phi)} \; ,   
\label{b2}
\eeqa
which are derived as the Euler-Lagrange equations of 
the Lagrangian (\ref{boh}). 

Assuming that both $\phi(t)$ and $z(t)$ are small, 
i.e. $|\phi(t)|\ll 1$ and $|z(t)|\ll 1$, 
the Lagrangian (\ref{boh}) can be approximated as 
\beq 
{L}_{ACS}^{(2)} = N \hbar\, z {\dot \phi} - 
{JN\over 2} \phi^2 - {(JN + U \left(1-{1\over N}\right) N^2)\over 2} z^2 \; ,  
\eeq 
removing a constant term. The Euler-Lagrange equations of this 
quadratic Lagrangian are the linearized Josephson-junction equations
\beqa 
\hbar \, \dot{\phi} &=& \left( J+UN \left(1-{1\over N}\right) \right) z  \; , 
\\
\hbar \, \dot{z} &=& - J\phi \; ,  
\eeqa
which can be rewritten as a single equation for the 
harmonic oscillation of $\phi(t)$ and the harmonic 
oscillation of $z(t)$, given by 
\beqa 
\ddot{\phi} + \Omega_{A}^2 \ \phi = 0 \; ,
\\
\ddot{z} + \Omega_{A}^2 \ z = 0 \; , 
\eeqa
both with frequency 
\beq 
\Omega_{A} = {1\over \hbar} 
\sqrt{J^2 + N U J \left(1-{1\over N}\right)} \; , 
\label{f-bmf}
\eeq
that is the atomic-coherent-state frequency of macroscopic quantum 
oscillation in terms of tunneling energy $J$, interaction strength $U$, 
and number $N$ of particles. Quite remarkably, this frequency is 
different and smaller with respect to the standard mean-field one, 
given by Eq. (\ref{frequenza-vera-mf}). However, the familiar 
mean-field result is recovered in the limit of a large number $N$ 
of bosonic particles. In addition, for $N=1$, Eq. (\ref{f-bmf}) 
gives $\Omega_A=J/\hbar$ that is the exact Rabi frequency of the 
one-particle tunneling dynamics in a double-well potential. 

In the same fashion as in the previous section, the conserved 
energy associated to Eqs. (\ref{b1}) and (\ref{b2}) reads 
\beq
{E}_{ACS} =  {U N^2\over 2} \left(1 - {1\over N}\right) 
z^2 - J N \sqrt{1-z^2} \, \cos{(\phi)} 
\eeq
and using the condition (\ref{trappingcondition}) we get the inequality 
\beq
\Lambda > 
\Lambda_{MQST,A} = { {1+\sqrt{1-z^2(0)} \cos(\phi (0))} \over z(0)^2/2} 
{1\over \left(1-{1 \over {N} }  \right)} \; ,
\label{mqstacs}
\eeq
where $\Lambda_{MQST,A}$ is the atomic-coherent-state MQST critical 
parameter in terms of 
tunneling energy $J$, interaction strength $U$, and number $N$ of particles. 
Remarkably this value is bigger than the standard mean field one, given by 
Eq. (\ref{mqst}), which is recovered in the semiclassical approximation 
of a large number $N$ of bosonic particles. 

In addition to the usual {\cred ground-state} 
stationary solutions (\ref{symmetricplus}) 
and (\ref{symmetricminus}) we obtain from the system of Eq. (\ref{b1}) and 
(\ref{b2}) a correction to the symmetry-breaking ones
\beqa
z_{ ACS \pm}&=&\pm \sqrt{1-{{1} \over {\Lambda^2}}
\left(1-{1 \over {N} }  \right)^{-2}   }  
\\
\phi_n &=& {\cred 2\pi n} 
\eeqa
with $n \in \mathbb{Z}$ and $\Lambda=NU/J$. 
It follows that, within the approach based on the atomic coherent state, 
the critical strength for the SSB of the balanced ground state 
($z=0$, $\phi=0$) reads 
\beq 
\Lambda_{SSB,A} = - {1\over \left(1 - {1\over N}\right)} \; .
\label{ssb-acs}
\eeq
This means that for $\Lambda=UN/J<\Lambda_{SSB,A}=1/(1-1/N)$ 
the ground state is not balanced. Clearly, for $N\gg 1$ from 
Eq. (\ref{ssb-acs}) one gets Eq. (\ref{ssb-cs}), while for $N=1$ one finds 
$\Lambda_{SSB,A}=-\infty$: {\cred within the ACS approach} 
with only one boson the spontaneous symmetry breaking cannot be obtained. 

\section{Numerical results}

To test our analytical results we compare them with {\cred numerical} 
simulations. The initial many-body state $|\Psi(0)\rangle$ 
for the time-dependent numerical simulations 
is the coherent state $|ACS(0)\rangle$ from Eq. (\ref{acs}), with 
a given choice of $z(0)$ and $\phi(0)$. 
The time evolved many-body state is then formally obtained as 
\beq 
|\Psi(t)\rangle = e^{-i{\hat H}t/\hbar} \, |\Psi(0)\rangle \; , 
\eeq
with ${\hat H}$ given by Eq. (\ref{ham}).  

Knowing $|\Psi(t)\rangle$ the population 
imbalance at time $t$ is given by 
\beq 
z(t) = \langle \Psi(t) | {{\hat N}_1 - {\hat N}_2\over N} 
|\Psi(t) \rangle \; . 
\label{zorro-ex}
\eeq

\begin{figure}[t]
\centerline{\epsfig{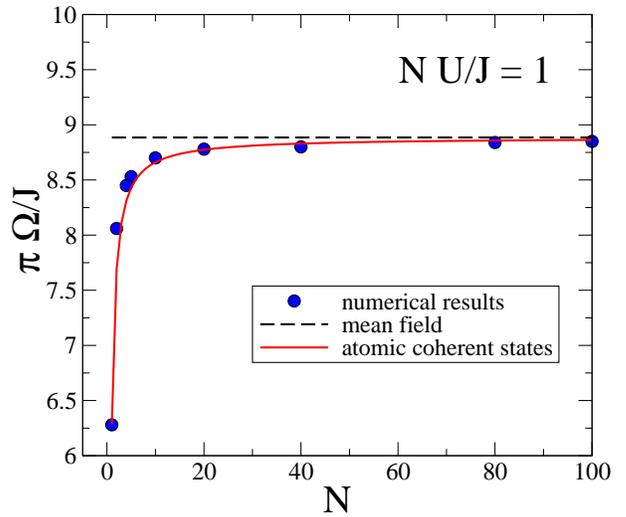}}
\small 
\caption{(Color online). Josephson frequency $\Omega$ as a function 
of the number $N$ of bosons, with $UN/J=1$, {\cred $J>0$,} and $\hbar=1$. 
Filled circles: {\cred numerical} results. Dashed line: 
mean-field result, Eq. (\ref{frequenza-vera-mf}), 
based on Glauber coherent states. Solid curve: results of 
Eq. (\ref{f-bmf}), based on atomic coherent states (ACS). 
Initial conditions: $z(0)=0.1$ and $\phi(0) =0$.} 
\label{fig1}
\end{figure} 

In Fig. \ref{fig1} we plot 
the Josephson frequency $\Omega$ as a function of the number $N$ 
of bosons, but with a fixed value of $UN/J=1$. 
As shown in the figure, the standard mean-field prediction (dashed 
curve), Eq. (\ref{frequenza-vera-mf}), predicts an horizontal line. 
The {\cred numerical} results (filled circles), which are very far from 
the standard mean-field predictions, are instead reproduced extremely well 
by Eq. (\ref{f-bmf}), based on atomic coherent states. 
Indeed, as previously stessed, for $N=1$ 
Eq. (\ref{f-bmf}) gives the correct Rabi frequency. 
{\cred However, this exact result is, in some sense, accidental  
since, as shown by the figure, for intermediate values of $N$ 
($4<N<10$) the agreement gets slightly worse.}

\begin{figure}[t]
\centerline{\epsfig{file=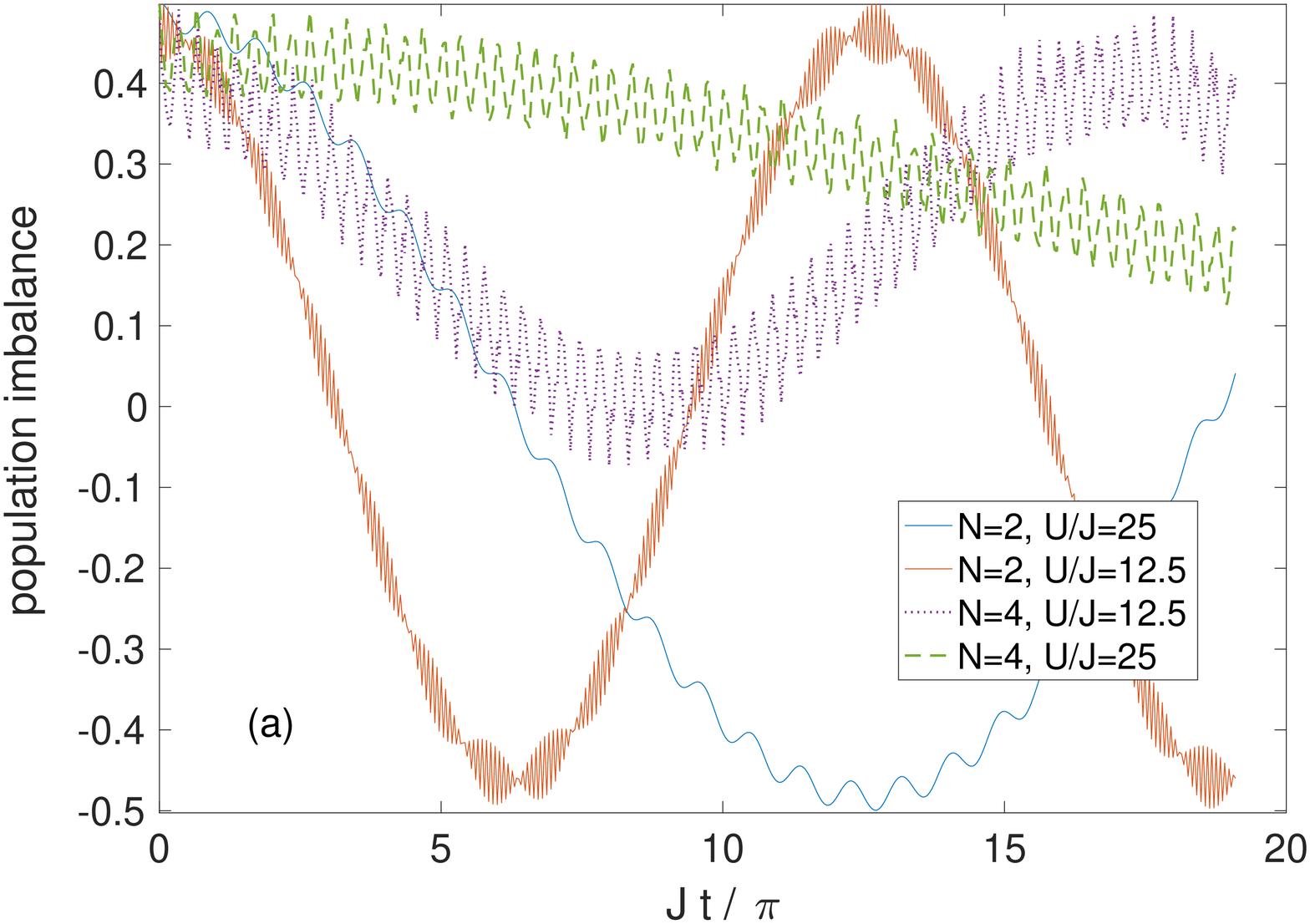,width=1\linewidth}}
\centerline{\epsfig{file=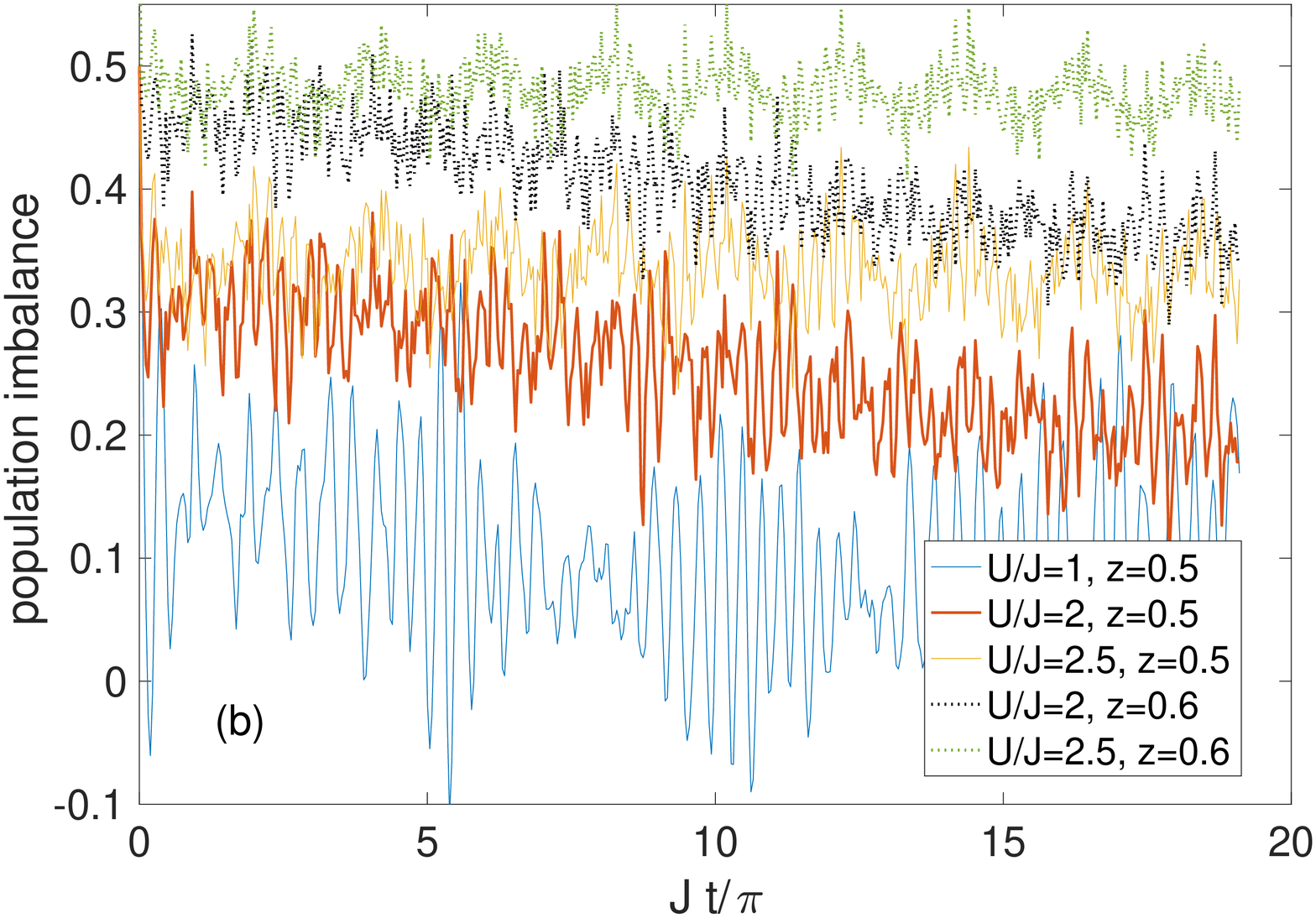,width=1\linewidth}}
\small 
\caption{(Color online). Time evolution of the {\cred numerical} 
population imbalance of Eq. (\ref{zorro-ex}) for different values of 
number \saw{$N=2$ and $4$ (a) and $N=20$ (b)
of bosons and interaction strength $U/J$, please see the legends, 
and} {\cred $J>0$.}
The initial quantum state $|ACS(0)\rangle$ 
is characterized by \saw{$\phi(0)=0$ and $z(0)=0.5$ (a, b) and 
$z(0)=0.6$ (only in b). Both panels highlight the difficulty in determining 
a critical value for self-trapping due to a smooth transitions 
to the oscillating regime (b) and the possibly long oscillation 
periods (a, b). Strict self-trapping seems to be absent for too 
small $N<5$ (a)}.} 
\label{fig:2}
\end{figure} 

We investigate numerically also the onset of macroscopic quantum 
self trapping (MQST). In Fig. \ref{fig:2} we report the {\cred numerical}  
time evolution 
of the population imbalance $z_{ex}(t)$ for different values of 
the number $N$ of bosons and of the interaction strength $NU/J$. 
In the figure the numerical results are obtained with an initial 
ACS state $|ACS(0)\rangle$ where $z(0)=0.5$ and $\phi(0)=0$. 
In general, during the time evolution the many-body quantum 
state $|\Psi(t)\rangle$ does not remains close to an atomic 
coherent state. This is especially true in the so-called Fock regime, 
where $U/J\gg N$ \cite{leggett2001}. Unfortunately, this is 
the regime where the MQST can be achieved. Fig. \ref{fig:2} 
illustrates the problems in determining a critical value for MQST:
for $N \lesssim 10$ interwell oscillations possibly occur with a
very long period even for very large values of $\Lambda$, 
\saw{see Fig. \ref{fig:2}(a)}. 
For larger $N=10, \dots , 100$, MQST is found, yet the loss of it 
occurs smoothly diminishing the interaction parameter, making it hard 
to define a 
critical value. \saw{Fig. \ref{fig:2}(b) illustrates this problem for 
$N=20$ and various values of $U/J$ 
for two slightly different initial conditions.} We opted for the definition
that just no crossing of zero imbalance should happen. 
This definition typically underestimates the values obtained from, e.g., 
mean-field theory, as seen in the next Fig. \ref{fig3}.

\begin{figure}[t]
\centerline{\epsfig{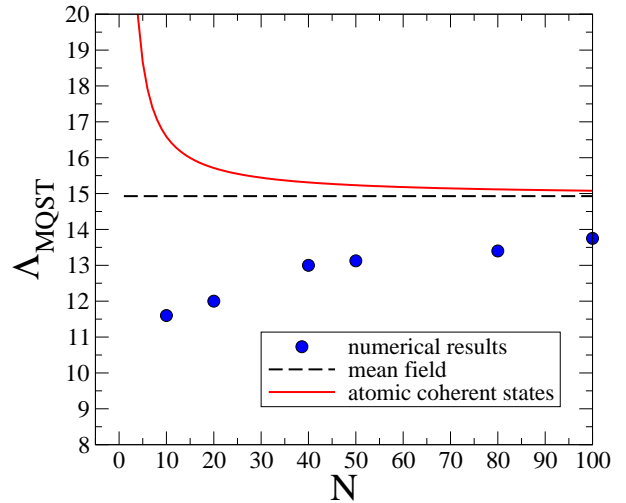}}
\small 
\caption{(Color online). Critical interaction strength $\Lambda_{MQST}$ 
for the macroscopic quantum self trapping (MQSF) 
as a function of the number $N$ of bosons. {\cred Notice that we take $J>0$}. 
Filled circles: {\cred numerical} results. Dashed line: 
mean-field result, Eq. (\ref{mqst}), 
based on Glauber coherent states. Solid curve: results of 
Eq. (\ref{mqstacs}), based on atomic coherent states (ACS). 
Initial conditions: $z(0)=0.5$ and $\phi(0)=0$.} 
\label{fig3}
\end{figure} 

In Fig. \ref{fig3} we show the critical interaction strength $\Lambda_{MQST}$ 
for the macroscopic quantum self trapping (MQSF) 
as a function of the number $N$ of bosons. 
In this case neither the mean-field results (dashed line) nor the 
ACS predictions (solid curve) are able to describe accurately 
the numerical findings (filled circles) for a small number $N$ 
of atoms. 

Let us conclude this Section by investigating 
the spontaneous symmetry breaking (SSB) of the ground state 
of the two-site Bose-Hubbard model, which appears for $U<0$ 
above a critical threshold \cite{sala-mazza1}. 
The exact number-conserving ground state of our system 
can be written as 
\beq 
|GS\rangle = \sum_{j=0}^N c_j \, |j\rangle_1 \otimes |N-j\rangle_2 \; , 
\eeq
where $|c_j|^2$ is the probability of finding the ground state 
with $j$ bosons in the site $1$ and $N-j$ bosons in the site $2$. 
Here $|j\rangle_1$ is the Fock state with $j$ bosons in the site $1$ 
and $|N-j\rangle$ is the Fock state with $N-j$ bosons in the site $2$. 
The amplitude probabilities $c_j$ are determined numerically by diagonalizing 
the $(N+1)\times (N+1)$ Hamiltonian matrix obtained from (\ref{ham}). 
Clearly these amplitude probabilities $c_j$ strongly depend 
on the values of the hopping 
parameter $J$, on-site interaction strength $U$, and total 
number $N$ of bosons. For $U>0$ 
the distribution ${\cal P}(|c_j|^2)$ of the probabilities $|c_j|^2$ 
is unimodal with its maximum at $|c_{N/2}|^2$ (if $N$ is even) 
\cite{sala-mazza1}. However, 
for $U<0$ the distribution ${\cal P}(|c_j|^2)$ becomes 
bimodal with a local minimum at $|c_{N/2}|^2$ (if $N$ is even) 
when $|U|$ exceeds a critical threshold \cite{sala-mazza1}. 

\begin{figure}[t]
\vspace{4mm}
\centerline{\epsfig{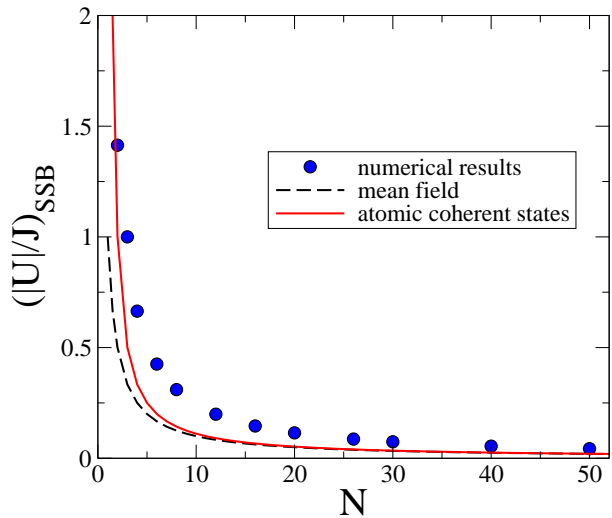}}
\small 
\caption{(Color online). Dimensionless interaction strength 
$(|U|/J)_{SSB}$ for the onset of spontaneous symmetry breaking (SSB)  
as a function of the number $N$ of bosons. {\cred Notice that we use $J>0$}. 
Filled circles: {\cred numerical} results 
obtained from the onset of a bimodal structure in the 
distribution ${\cal P}(|c_j|^2)$. 
Dashed line: mean-field result $(|U|/J)_{SSB}=1/N$ 
based on Glauber coherent states, see Eq. (\ref{ssb-cs}). Solid curve: 
$(|U|/J)_{SSB}=1/(N-1)$, based on atomic coherent states (ACS), 
see Eq. (\ref{ssb-acs}).} 
\label{fig4}
\end{figure} 

The semiclassical SSB, described by Eqs. (\ref{ssb-cs}) 
and (\ref{ssb-acs}), corresponds 
in a full quantum mechanical treatment to the onset 
of the bimodal structure in the distribution ${\cal P}(|c_j|)$ 
\cite{sala-mazza1}. 
In Fig. \ref{fig4} we report the dimensionless interaction strength 
$(|U|/J)_{SSB}$ for the spontaneous symmetry breaking (SSB)  
as a function of the number $N$ of bosons. 
In the figure we compare the {\cred numerical} results (filled 
circles)  \cite{sala-mazza1} with the semiclassical 
predictions based on Glauber coherent states (dashed curve) 
and atomic coherent states (solid curve). 
The figure shows that the {\cred numerical} results of 
SSB are quite well approximated by the ACS variational approach, 
which is more accurate with respect to the standard mean-field one. 
For large $N$ the {\cred numerical} results end up in the 
analytical curves, which become practically indistinguishable. 

\section{Conclusions}

In this paper we have adopted a second-quantization formalism 
and time-dependent atomic coherent states to study finite-size 
effects in a Josephson junction of $N$ bosons, obtaining 
experimentally detectable theoretical predictions. 
The experiments with cold atoms in lattices and double wells 
reported in \cite{Ober1,Ober2,Ober3,Ober4}, for instance, 
showed that atom numbers well below $N=100$ can be reached and 
successfully detected with an uncertainty of the order one atom. 
In particular we have obtained an analytical formula 
with $1/N$ corrections to the standard mean-field treatment 
for the frequency of Josephson oscillations. 
We have shown that this formula, based on atomic coherent states,  
is in very good agreement with {\cred numerical} simulations 
and it reduces to the familiar mean-field one in the large $N$ limit. 
We have also investigated the spontaneous symmetry 
breaking of the ground state. At the 
critical interaction strength for the spontaneous symmetry breaking  
the population-balanced configuration is no more the one with 
maximal probability. 
Also in this case the agreement between the analytical predictions 
of the atomic coherent states and {\cred numerical} results is good. 
Finally, we have studied the critical interaction strength 
for the macroscopic quantum self trapping. Here we have found 
that the $1/N$ corrections to the standard mean-field theory 
predicted by the atomic coherent states do not work. 
Summarizing, the time-dependent variational 
ansatz with atomic coherent states is quite reliable 
in the description of the short-time dynamics of the bosonic 
Josephson junction both in the Rabi regime, 
where {\cred $0\leq |U/J| \ll 1/N$}, and in the Josephson 
regime, where {\cred $1/N \ll |U/J| \ll N$} \cite{leggett2001}. 
Instead, in the Fock regime, where {\cred $|U/J| \gg N$}, a full many-body 
quantum treatment is needed. 

\section*{Acknowledgements}

The authors thank A. Cappellaro, L. Dell'Anna, A. Notari, 
V. Penna, and F. Toigo for useful discussions. 
LS acknowledges the BIRD project ``Superfluid properties of Fermi gases 
in optical potentials'' of the University of Padova for financial support.

\end{document}